\documentclass [12pt] {article}
\usepackage{amsfonts}
\usepackage{amsmath}
\usepackage{graphics}
\usepackage{epsfig}

\begin{document}

\def\alt{\mathrel{\hbox{\rlap{\hbox{\lower4pt\hbox{$\sim$}}}\hbox{$<$}}}}
\def\agt{\mathrel{\hbox{\rlap{\hbox{\lower4pt\hbox{$\sim$}}}\hbox{$>$}}}}

\thispagestyle{empty}

\begin{flushright}
hep-th/0310075\\ October 2003
\end{flushright}

\noindent
\vskip3.3cm
\begin{center}
{\Large\bf  Vacuum Polarization in an Anti-de~Sitter Space}
\end{center}

\vspace{-0.7cm}

\begin{center}
{\Large\bf  as an Origin for a Cosmological Constant}
\end{center}

\vspace{-0.7cm}

\begin{center}
{\Large\bf  in a Brane World}
\end{center}
\bigskip\bigskip\bigskip

\begin{center}
{\large Li-Xin Li}
\medskip

{\em Harvard-Smithsonian Center for Astrophysics, Cambridge,
     MA 02138, USA}\\
\medskip
{\small\tt lli@cfa.harvard.edu} \\
\bigskip
\end{center}

\bigskip 
\begin{center}
{\sc Abstract}
\end{center}
\noindent
In this Letter we show that the vacuum polarization of quantum fields
in an anti-de~Sitter space
naturally gives rise to a small but nonzero cosmological constant in a
brane world living in it. To explain the extremely small ratio of mass
density in the cosmological constant to the Planck mass density in our
universe ($\approx 10^{-123}$) as suggested by cosmological observations,
all we need is a four-dimensional brane world (our universe) living in
a five-dimensional anti-de~Sitter space with a curvature radius $r_0\sim 
10^{-3}$cm and a fundamental Planck energy $M_{\rm P}\sim 10^9$GeV, and a 
scalar field with a mass $m\sim r_0^{-1}\sim 10^{-2}$eV. Probing gravity
down to a scale $\sim 10^{-3}$cm, which is attainable in the near future, 
will provide a test of the model.

\vfill
\setcounter{page}{0}
\setcounter{footnote}{0}

\newpage



Much attention has recently been paid to the idea that our universe is a
four-dimensional brane (three dimensions of space plus one dimension of time) 
embedded in a five-dimensional bulk universe. This fact is mainly caused by the 
discoveries that by confining the standard model particles on a brane the 
extra dimensions can be larger than previously anticipated then the very large 
hierarchy between the electroweak and the Planck energy scales is relaxed 
\cite{ant98,ark98}, and that on a brane embedded in a five-dimensional 
anti-de~Sitter space Newton's gravitational law can be recovered on distances 
larger than submillimeter as required by the gravity experiments that have
been taken so far \cite{ran99a,ran99b,gar00}. A lot of work has since then been 
done on brane cosmology, and people have shown that in the case when the bulk
universe is empty away from the brane the cosmology on the brane deviates from 
that given by the standard general relativity only in the epoch before the 
cosmic nucleosynthesis when the energy density in the universe is sufficiently 
high (\cite{ida99,kra99,bin00,sto00,bow00,kho01,lan02,bra03} and references 
therein). Observational test of brane models by supernovae has also been 
discussed \cite{dab02}. 

Despite the remarkable success of the brane model, the cosmological constant 
problem remains unsolved \cite{bra03}. In the brane scenario, it is usually 
assumed that the tension of the brane exactly cancels the effect of the
negative cosmological constant in the bulk universe, giving rise to a zero
cosmological constant in the brane universe. However, observations of
gravitational lensing, high redshift type Ia supernova, cosmic microwave
background, and cluster counts consistently show that the universe is in
all likelyhood flat and accelerating, with the present mass density being
composed of about 70\% of cosmological constant (or dark energy), 30\% of dark
and ordinary matter \cite{ost95,bah99,wan00,pea01,efs02,yam02,spe03}.
In Planck units $\hbar = c = 1$, the cosmological constant is extremely 
small: $\rho_\Lambda/\rho_{\rm P} \approx 10^{-123}$, where $\rho_\Lambda$
is the mass density in the cosmological constant, $\rho_{\rm P}$ is the 
Planck mass density in our four-dimensional universe. The existence of such
a small but nonzero cosmological constant is usually called the cosmological
constant problem and continues to be a big mystery for modern cosmology and
theories of elementary particles \cite{wei89,pee03,pad03}. For current 
efforts toward solving this problem, see 
\cite{ark00,bra01,gar01,fla01,gar02,gup02,mof02,cor03,gue03,kim03,igl03,set03} 
and references therein.

In this Letter we propose an explanation for the existence of a small but 
nonzero cosmological constant in the brane world scenario. For simplicity, we
assume that the bulk spacetime is a five-dimensional anti-de~Sitter space. 
We show 
that the vacuum polarization of a massive scalar field in the bulk spacetime
has a stress-energy tensor that behaves like a small positive cosmological 
constant. As in the standard scenario of a brane world we assume that the
brane tension cancels the effect of the bare cosmological constant in the
anti-de~Sitter space, then on the brane a small but nonzero cosmological
constant exists, which we will show is consistent with cosmological observations
if the mass of the scalar field is $\sim r_0^{-1}$ where $r_0$ is the 
curvature radius of the anti-de~Sitter space.

An anti-de~Sitter space is a spacetime that has the maximal symmetry and a 
negative constant curvature, supported by a negative cosmological constant 
\cite{haw73,li99}. For an $N$-dimensional anti-de~Sitter space, its curvature
radius is related to the cosmological constant by $r_0 =\left[(N-1)(N-2)/
(-2\Lambda)\right]^{1/2}$. The Riemann curvature tensor is $R_{abcd} =2
\Lambda(g_{ac}\,g_{bd} - g_{ad}\,g_{bc})/[(N-1)(N-2)]$.
The corresponding Ricci tensor $R_{ab} = 2\Lambda g_{ab}/(N-2)$ satisfies the 
$N$-dimensional Einstein equation $R_{ab} -(1/2)\, R\, g_{ab} + \Lambda\, 
g_{ab}= 8\pi M_{\rm P}^{2-N}\, T_{ab}$ when the $N$-dimensional 
stress-energy tensor $T_{ab} = 0$, where $R = g^{ab} R_{ab} = 2\Lambda
N/(N-2)$, $M_{\rm P}$ is the $N$-dimensional Planck mass.

Quantum fields in an anti-de~Sitter space have been studied extensively
\cite{avi78,dul85,bur85,all86,cam91,cam92,kam99,cal99,li99}. For a quantum
state that is maximally symmetric (i.e., anti-de~Sitter-invariant), the
two-point function $G(x,x^\prime)$ 
depends only on the geodesic distance $\mu(x,x^\prime)$ between $x$ and 
$x^\prime$. For a massive scalar field $\phi(x)$ in an $N$-dimensional 
anti-de~Sitter space of radius $r_0$, the two-point function for a maximally 
symmetric vacuum state, $G(x,x^\prime) = \langle 0|\phi(x)\phi(x^\prime)
|0\rangle$, is given by \cite{all86}
\begin{eqnarray}
     G(x,x^\prime) = G(\mu) =A \cosh^{-2\alpha}\left(\frac{\mu}{2 r_0}\right)\,
	     F\left(\alpha,\alpha-\frac{N}{2}+1; 2\alpha -N +2; \cosh^{-2}
		\frac{\mu}{2 r_0}\right) \;,
	\label{gxx}
\end{eqnarray}
where 
\begin{eqnarray}
     \alpha\equiv \frac{1}{2}\left[N-1+\sqrt{(N-1)^2+4m^2 r_0^2}\right] \;,
	     \hspace{1cm}
     A\equiv \frac{\Gamma(\alpha)\,r_0^{2-N}}
	     {2^{2\alpha+1} \pi^{(N-1)/2} \Gamma\left(\alpha-\frac{N}{2}
		+\frac{3}{2}\right)} \;.
	\label{alp}
\end{eqnarray}
Here $F$ denotes the hypergeometric function, $\Gamma$ denotes the Gamma
function, and $m$ is the mass of the scalar field particle.

The two-point function given by Eq.~(\ref{gxx}) satisfies the following
two conditions: (a) falls off as fast as possible at spatial infinity 
$\mu^2\rightarrow\infty$, and (b) has the same strength $\mu^2\rightarrow
0$ singularity as in a flat space. It is worth to note that the two-point
function so defined can be obtained from the continuation of a Euclidean 
Green function on $H^N$---an $N$-dimensional hyperbolic Euclidean space, but
cannot be obtained from the continuation of a Euclidean Green function on 
$S^N$---an $N$-dimensional sphere \cite{bur85,cam91}.

For $N<2$ the limit $\mu\rightarrow 0$ in Eq.~(\ref{gxx}) is finite, and we
have
\begin{eqnarray}
     G(x,x) = \frac{\Gamma\left(1-\frac{1}{2}N\right)\Gamma(\alpha)}{2^N 
	     \pi^{N/2}\Gamma\left(\alpha-N+2\right)}\; r_0^{2-N} \;.
	\label{gx2}
\end{eqnarray}
$G(x,x)$ is analytic in the complex $N$-plane apart from simple poles. 
Therefore, it may be extended throughout the whole complex $N$-plane 
\cite{can75} (see also \cite{bur85,kam99}). Especially, when $N$ is odd, the
$G(x,x)$ given by Eq.~(\ref{gx2}) is finite and can be taken to be the 
regularized value for $\phi(x)^2$, according to the dimensional regularization 
procedure \cite{bir82}.

For a vacuum state that is maximally symmetric the expectation value of its 
stress-energy tensor must be given by $\langle T_{ab}\rangle =T\,g_{ab}/N$, 
where $T = g^{ab}\langle T_{ab}\rangle$ is the trace \cite{bir82}. For
a massive scalar field whose $G(x,x)$ is given by Eq.~(\ref{gx2}), which 
is finite and constant when $N$ is odd, we have $T = - m^2\langle\phi(x)^2 
\rangle= - m^2 G(x,x)$. Then we have
\begin{eqnarray}
     \langle T_{ab}\rangle = -\frac{\Gamma\left(1-\frac{1}{2}N\right)
	     \Gamma(\alpha)}{N 2^N \pi^{N/2}\Gamma\left(\alpha-N+2\right)}
	     \; m^2 r_0^{2-N} g_{ab}\;,
	\label{tab}
\end{eqnarray}
which is the renormalized stress-energy tensor when $N$ is odd 
\cite{cal99}.\footnote{In this Letter we focus on the case when $N$ is odd 
since we have assumed the bulk space has five dimensions. When $N$ is even, 
further care must be taken to remove the poles in the Gamma function, 
see \cite{can75,bir82,bur85,kam99}.}
The stress-energy tensor in Eq.~(\ref{tab}), which describes the vacuum 
polarization effect of a massive scalar field in an odd $N$-dimensional 
anti-de~Sitter space, corresponds to a cosmological constant
\begin{eqnarray}
     \Lambda^\prime = \frac{\Gamma\left(1-\frac{1}{2}N\right)
	     \Gamma(\alpha)}{N 2^{N-3} \pi^{(N-2)/2}\Gamma\left(\alpha-
	     N+2\right)}\; m^2 \left(M_{\rm P}r_0\right)^{2-N}\;.
	\label{lamp}
\end{eqnarray}

When $N = 5$---the case that is of particular interest in this Letter, we have
\begin{eqnarray}
     \Lambda^\prime = \frac{m^2}{15 \pi M_{\rm P}^{3} r_0^3}\left(3 +
	     m^2 r_0^2\right) \sqrt{4+ m^2 r_0^2} \;,
	\label{lamp5}
\end{eqnarray}
which is positive when $m^2 r_0^2 >0$ or $-4<m^2 r_0^2 <-3$, negative when 
$-3<m^2 r_0^2 <0$, zero when $m^2 r_0^2 = 0, -3$, or $-4$.

The gravitational equation on a brane in the five-dimensional anti-de~Sitter
space is then (see, e.g., \cite{shi00})
\begin{eqnarray}
     {}^{(4)}R_{ab} - \frac{1}{2}\,{}^{(4)}R\,{}^{(4)}g_{ab} +
	     \frac{1}{2}\left(\Lambda+\Lambda^\prime\right)\,{}^{(4)}g_{ab}
		&=& -\frac{16\pi^2}{3} M_{\rm P}^{-6}\sigma^2\, {}^{(4)}g_{ab} +
		\frac{32\pi^2}{3} M_{\rm P}^{-6} \sigma\, {}^{(4)}T_{ab} \nonumber\\
		&&+ \left[\mbox{higher order terms}\right] \;,
\end{eqnarray}
where the index ``(4)'' on the left shoulder denote a quantity defined on the
four-dimensional brane world, $\sigma$ is the tension of the brane, 
${}^{(4)}T_{ab}$ is the 
stress-energy tensor of matter (with the tension term subtracted) on the brane,
and ``higher order terms'' stands for terms quadratic in ${}^{(4)}T_{ab}$ that
are important only in sufficiently high energy regime. Following the standard
procedure in the brane world scenario, let us set
\begin{eqnarray}
     -\frac{16\pi^2}{3} M_{\rm P}^{-6}\sigma^2 = \frac{1}{2} \Lambda \;,
	\label{lam40} \hspace{1cm}
     \frac{32\pi^2}{3} M_{\rm P}^{-6} \sigma = 8\pi m_{\rm P}^{-2} \;,
	\label{g4}
\end{eqnarray}
where $m_{\rm P} = G^{-1/2}$ is the Planck mass in the four-dimensional space.
Then, the standard Einstein equation (without a cosmological constant) is
recovered if $\Lambda^\prime =0$ and higher orders terms are neglected. By
the second equation of (\ref{g4}), $\sigma$ must be positive, i.e. the brane 
must have a positive tension.

If $\Lambda^\prime \neq0$ and is given by Eq.~(\ref{lamp5}), then there is a
cosmological constant in the brane world, whose value is
\begin{eqnarray}
     {}^{(4)}\Lambda = \frac{m^2}{30 \pi M_{\rm P}^{3} r_0^3}\left(3 +
	     m^2 r_0^2\right) \sqrt{4+ m^2 r_0^2} \;.
	\label{lam4}
\end{eqnarray}
Of course, one can argue that Eq.~(\ref{lam40}) can be modified to cancel
$\Lambda^\prime$ then the cosmological constant in the brane world will be
zero. However, doing so would introduce more fine tune to the
theory as Eq.~(\ref{lam40}) is already a fine tune. We think the true question
is, for reasonable values of parameters, if the ${}^{(4)}\Lambda$ given by
Eq.~(\ref{lam4}) is consistent with the value of a cosmological constant that
has been implied by the cosmological observations. In the rest of the Letter 
we will show that the answer to this question is ``Yes''. 

By construction, there are two fundamental energy scales in the five-dimensional 
anti-de~Sitter bulk space: $M_{\rm P}$ and $r_0^{-1}$. The four-dimensional 
Planck mass $m_{\rm P}$ and the brane tension $\sigma$ are derived from $M_{\rm 
P}$ and $r_0^{-1}$ by Eq.~(\ref{lam40}):
\begin{eqnarray}
     m_{\rm P}^2 = M_{\rm P}^3 r_0 \;, \hspace{1cm}
	\sigma = \frac{3}{4\pi} M_{\rm P}^3 r_0^{-1} \;,
	\label{mps}
\end{eqnarray}
where $\Lambda = -6/r_0^2$ has been used. The requirement to recover
Newton's gravitational law down to scales of the submillimeter order (see 
\cite{hoy01,lon03}) puts a stringent constraint on $r_0$: $r_0 < 10^{-2}$cm, 
which by Eq.~(\ref{mps}) implies that $M_{\rm P} >5.4\times 10^{-11} m_{\rm P} 
\approx 6.7\times 10^8$GeV and $\sigma > \left(2.8\times 10^{-16} m_{\rm P}
\right)^4 \approx \left(3000{\rm GeV}\right)^4$.

The four-dimensional cosmological constant ${}^{(4)}\Lambda$ depends on another
parameter: the mass of the scalar field, $m$, which in principle should be
constructed from $M_{\rm P}$ and/or $r_0^{-1}$. Since $M_{\rm P} \gg r_0^{-1}$
[otherwise Eq.~(\ref{mps}) would imply $m_{\rm P} \alt M_{\rm P}$], a natural
possibility would be $m \sim r_0^{-1}$, not $m \sim M_{\rm P}$. 

An elegant argument for a discrete mass spectrum for a scalar field in an
anti-de~Sitter space was given by Allen and Jacobson \cite{all86}. In an
anti-de~Sitter space, which has topology $S^1\times {\mathbb R}^{N-1}$ and 
contains closed timelike curves, the two-point function
should be invariant under the transformation $\mu\rightarrow \mu + i\,2\pi r_0 
j$,\footnote{Note, when $x$ and $x^\prime$ are timelikely separated $\mu$ is
an imaginary number.}
where $2\pi r_0$ is the timelike circumference at the ``neck'' 
of the anti-de~Sitter space, $j$ is any integer. From Eq.~(\ref{gxx}) we have
\begin{eqnarray}
     G(\mu+i\, 2\pi r_0 j) = \exp(-i\,2\pi j\alpha)\, G(\mu) \;,
	\label{gext}
\end{eqnarray}
then $\alpha$ must be an integer so that $G(\mu+i\, 2\pi r_0 j) = G(\mu)$. Let
$\alpha = n+4$, where $n$ is an integer, then from Eq.~(\ref{alp}) we have
\begin{eqnarray}
     m^2 = \frac{n(n+4)}{r_0^2} \;, \label{m2}
\end{eqnarray}
where we have set $N = 5$. When $n = 0, -1, -2, -3$, or $-4$, $m^2 \leq 0$ and 
from Eq.~(\ref{lam4}) we have ${}^{(4)}\Lambda =0$. When $n \ge 1$ or $n\le 
-5$, $m^2 >0$ and from Eq.~(\ref{lam4}) we have ${}^{(4)}\Lambda >0$. Since
$m^2 <0$ leads to the existence of tachyons which are usually thought 
unphysical, we restrict our treatment to the case of $m^2\ge 0$ and assume
$n = 0, 1, 2, ...$ in Eq.~(\ref{m2}). [Since Eq.~(\ref{m2}) is invariant 
under the transformation $n\rightarrow -n-4$, $n = -4, -5, -6, ...$ give same
values of $m^2$ as $n = 0, 1, 2, ...$.] The above argument supports the 
assumption that $r_0^{-1}$ is the natural scale for the mass $m$.

Of course, one can avoid closed timelike curves by ``unwrapping'' the $S^1$ 
dimension, or equivalently, considering the universal covering space of an
anti-de~Sitter space with topology ${\mathbb R}^N$. Then, the mass spectrum 
of the scalar field will be continuous and Eq.~(\ref{gext}) gives an explicit 
expression for $G(x,x^\prime)$ when $x$ and $x^\prime$ are separated by $j$ 
sheets \cite{all86}. However, we point out that in the brane world scenario
it is not necessary to go to the universal covering space to avoid closed
timelike curves in the brane space where we live in. Since all ordinary 
matter is assumed to be trapped on the brane, ordinary particles and 
photons cannot leave the brane to enter orbits with closed timelike or 
null curves. So, the closed timelike 
curves in the bulk anti-de~Sitter space will not bother us. (Certainly 
gravitons can leave the brane and enter the region with closed timelike 
curves.)

From Eq.~(\ref{lam4}) we obtain the ratio of the mass density in the 
cosmological constant to the Planck mass density in the brane universe
\begin{eqnarray}
     \frac{\rho_\Lambda}{\rho_{\rm P}} = \frac{f(mr_0)}{40 \pi^2}
		\left(M_{\rm P} r_0\right)^{-6}
	     = \frac{f(mr_0)}{40 \pi^2} \left(\frac{M_{\rm P}}{m_{\rm P}}
		\right)^{12} \;,
	\label{rlp}
\end{eqnarray}
where $\rho_\Lambda = {}^{(4)}\Lambda/8\pi m_{\rm P}^{-2}$, $\rho_{\rm P} = 
m_{\rm P}^4$, and $f(x)\equiv x^2 \left(1+x^2/3\right)\sqrt{1+x^2/4}$\,.
Because of the high power dependence of $\rho_\Lambda/\rho_{\rm P}$ on 
$M_{\rm P}/m_{\rm P}$ [$\rho_\Lambda/\rho_{\rm P}\propto \left(M_{\rm P}/
m_{\rm P}\right)^{12}$], an extremely small ratio of $\rho_\Lambda/\rho_{\rm 
P}$ is easily generated if $M_{\rm P}/m_{\rm P} \ll 1$. For example, if $M_{\rm 
P}/m_{\rm P} \sim 10^{-10}$ and $f(mr_0)\sim 1$, then $\rho_\Lambda/\rho_{\rm 
P}\sim 10^{-123}$, in agreement with the observation value.

\begin{figure}
\vspace{2.0cm}
\includegraphics[width=16cm]{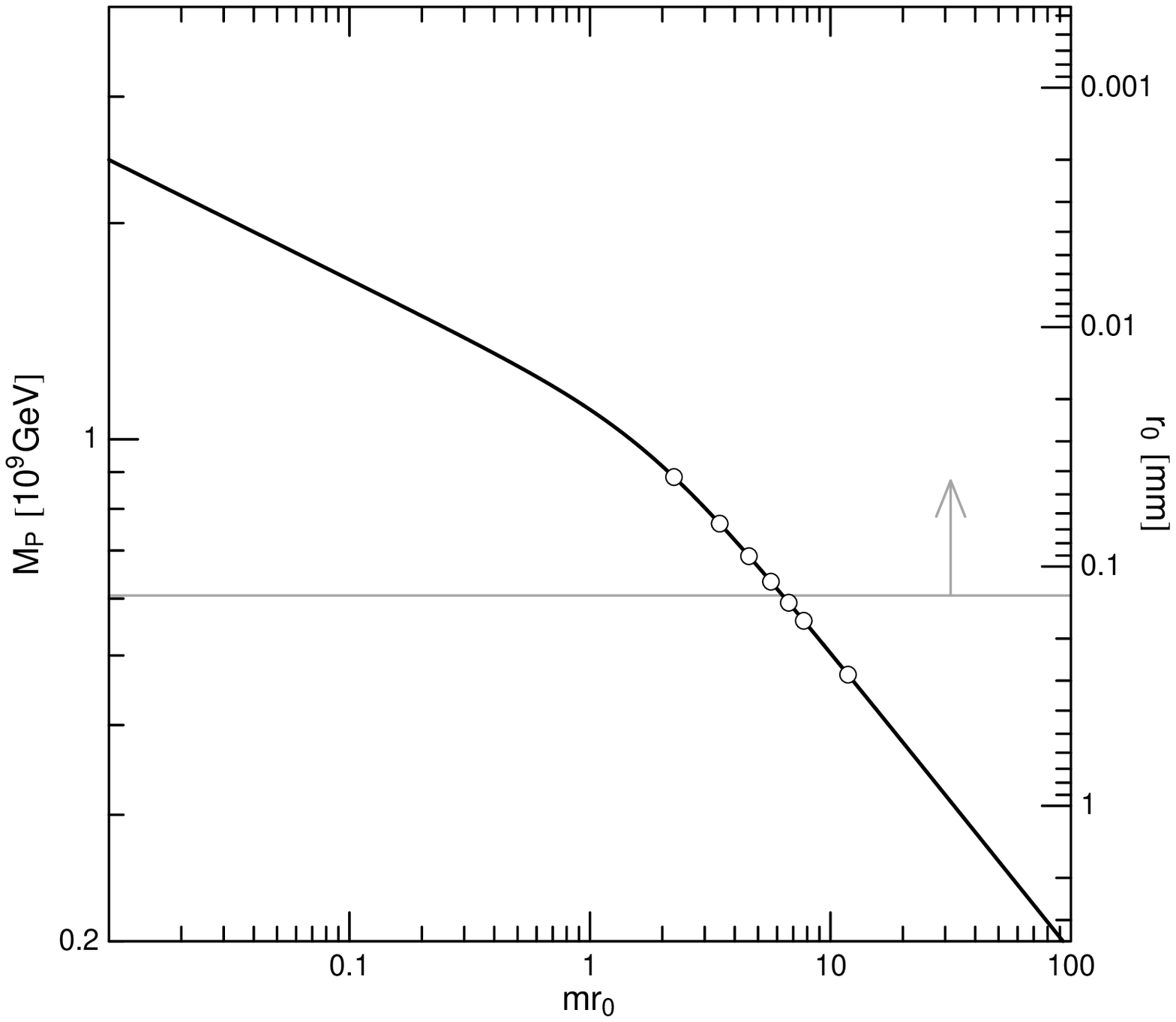}
\caption{Plot of $M_{\rm P}$ as a function of $m r_0$, by which the ratio 
$\rho_{\Lambda}/\rho_{\rm P} \approx 10^{-123}$ is produced on the brane 
universe [see Eq.~(\ref{rlp})]. The corresponding curvature radius of the bulk 
anti-de~Sitter space, $r_0$ [related to $M_{\rm P}$ by Eq.~(\ref{mps})], is 
indicated on the right vertical axis. The horizontal line is the limit given by
the gravity experiment in \cite{lon03}, with an upward arrow indicating that 
it is a lower bound. The circles on the curve show the mass $m$ given by 
Eq.~(\ref{m2}), for $n = 1, 2, 3, 4, 5, 6$, and $10$ (downward).
\label{fig1}}
\end{figure}

In Fig.~\ref{fig1}, we plot $M_{\rm P}$ as a function of $mr_0$, by setting 
$\rho_\Lambda/\rho_{\rm P} = 10^{-123}$, the value indicated by the observations
(assuming the Hubble constant $H_0 = 65\, {\rm km}\, {\rm s}^{-1}\, {\rm 
Mpc}^{-1}$). The limit given by the current gravity experiment \cite{lon03} 
is shown with a horizontal line, with an upward arrow indicating that is a 
lower bound. In \cite{lon03} no deviation from Newton's gravitational law 
was observed down to a distance of $d=108 \mu$m. We convert this distance to 
a limit on $r_0$ by $r_0 = (3/
2)^{1/2} d$ \cite{gar00}. The circles on the curve in Fig.~\ref{fig1} correspond 
to the mass given by Eq~(\ref{m2}), for $n = 1, 2, 3, 4, 5, 6$, and $10$ 
(downward). For the case with a discrete mass spectrum, which corresponds to an
anti-de~Sitter space with closed timelike curves, the limit given by \cite{lon03} 
excludes all solutions with $n\ge 5$. For the case with a continuous mass
spectrum, which corresponds to the universal covering space of an anti-de~Sitter
space without closed timelike curves, the limit given by \cite{lon03} excludes 
all solutions with $m > 6.3 /r_0$. In both cases the solutions being excluded 
would produce a too large cosmological constant if the limit from the gravity
experiment is satisfied. However, Fig.~\ref{fig1} shows that, there exist a range
of solutions that are within the limit given by the gravity experiment and
give rise to a cosmological constant with $\rho_{\Lambda}/\rho_{\rm P} \approx
10^{-123}$ in the brane universe. These solutions correspond to $n = 1, 2, 3$, or
$4$ for the case with a discrete mass spectrum, or $m < 6.3 /r_0$ for the case
with a continuous mass spectrum. 

To conclude, we have shown that the vacuum polarization effect of quantum 
fields in an anti-de~Sitter space naturally gives rise to a small but nonzero 
cosmological constant in a brane universe living in the anti-de~Sitter space. 
If our four-dimensional universe is embedded in a five-dimensional 
anti-de~Sitter space with a curvature radius $r_0 \sim 10^{-3}$cm and a 
fundamental Planck mass $M_{\rm P}\sim 10^9$GeV, then the stress-energy tensor 
arising from the vacuum polarization of a scalar field with a mass $m\sim 
r_0^{-1}\sim 10^{-2}$eV in the bulk anti-de~Sitter space behaves like a 
cosmological constant in our brane universe with a value in agreement with that 
suggested by current cosmological observations. Probing gravity down to a 
scale $\sim 10^{-3}$cm (which is becoming possible \cite{lon03,chi03}) will 
provide a test for the model presented in this Letter.


\clearpage

\end{document}